\documentclass[
    ,final            
    ,sort&compress    
  ]
  {aipproc}

\layoutstyle{6x9}
   \usepackage{graphicx}
   \usepackage{epstopdf}
  \usepackage{amssymb}
   \usepackage{pdfsync}
   \usepackage{color}

\def\teq#1{$\, #1\,$}                         
\def\apj{ApJ}
\def\apjsupp{ApJ Supp.}

\def\asr{Adv. Space Res.}                       

\def\mnras{{M.N.R.A.S.}}
\def\rmp{Rev. Mod. Phys.}                       
\newcommand{\vol}[2]{$\,$\rm #1\rm , #2.}           
\font\fiverm=cmr5             \font\sevenrm=cmr7

          \font\sixrm=cmr6       

\def\dover#1#2{\hbox{${{\displaystyle#1 \vphantom{(} }\over{
   \displaystyle #2 \vphantom{(} }}$}}
\def\erg{\varepsilon}     
\def\lambar{\lambda\llap {--}_{\rm c}} 
\def\fsc{\alpha_{\sevenrm f}}      
\def\epspar{\epsilon_{\parallel}}                          
\def\rL{r_{\hbox{\sixrm L}}}
\def\rns{R_{\hbox{\sixrm NS}}}
\def\rlc{R_{\hbox{\sixrm LC}}}
\def\Blc{B_{\hbox{\sixrm LC}}}

\def\gammax{\gamma_{\hbox{\sixrm MAX}}}
\def\gammadotacc{\dot{\gamma}_{\hbox{\sixrm acc}}}
\def\gammadotCR{\dot{\gamma}_{\hbox{\sixrm CR}}}
\def\gammadotSR{\dot{\gamma}_{\hbox{\sevenrm syn}}}

\def\Emax{E_{\hbox{\sixrm MAX}}} 
\def\emax{\erg_{\hbox{\sixrm MAX}}} 
\def\eesc{\erg_{\hbox{\sevenrm esc}}}      
\def\sigmaIG{\sigma_{\hbox{\sixrm IG}}}
\def\sigmaOG{\sigma_{\hbox{\sixrm OG}}}
\def\chiIG{\chi_{\hbox{\sixrm IG}}}

\begin{document}
\begin{flushright}
\phantom{p}
\vspace{-40pt}
     To appear in Proc. of the {\it Astrophysics of Neutron Stars 2010} Conference\\ 
     eds. E. G{\"o}{\u g}{\"u}{\c s}, T. Belloni \& U. Ertan (AIP Conf. Proc., New York).
\end{flushright} 

\title{Perspectives on Gamma-Ray Pulsar Emission}

\author{Matthew G. Baring}{
    address={Department of Physics and Astronomy, MS-108,
                      Rice University, P. O. Box 1892, \\
                      Houston, TX 77251-1892, USA \ \ {\rm Email: baring@rice.edu}}
}



\begin{abstract}
Pulsars are powerful sources of radiation across the electromagnetic
spectrum.  This paper highlights some theoretical insights into
non-thermal, magnetospheric pulsar gamma-ray radiation.   These advances
have been driven by NASA's {\it Fermi} mission, launched in mid-2008.
The Large Area Telescope (LAT) instrument on {\it Fermi} has afforded
the discrimination between polar cap and slot gap/outer gap acceleration
zones in young and middle-aged pulsars.  Altitude discernment using the
highest energy pulsar photons will be addressed, as will spectroscopic
interpretation of the primary radiation mechanism in the LAT band,
connecting to both polar cap/slot gap and outer gap scenarios.   Focuses
will mostly be on curvature radiation and magnetic pair creation,
including population trends that may afford probes of the magnetospheric
accelerating potential. 
\end{abstract}

\maketitle

\section{Introduction}
\label{sec:Introduction}
Prior to the launch of the Compton Gamma-Ray Observatory (CGRO) in 1991,
the gamma-ray pulsar database was extremely limited.  CGRO extended the
population to seven sources, all of them young pulsars, and only Geminga
had no definitive observation of a radio pulsar counterpart. There were
also a few marginal CGRO detections, including the millisecond pulsar
PSR J0218+4232. Following the launch of the {\it Fermi} Gamma-Ray Space
Telescope on June 11, 2008, the gamma-ray pulsar database has grown
rapidly, with over 75 sources now observed with high significance.  Most
of these are again young pulsars with higher spin-down luminosities than
the average for radio pulsars.  Yet, a sizable millisecond pulsar (MSP)
population has emerged, at present numbering more than twenty. Moreover,
the improved photon counts captured by the {\it Fermi} Large Area
Telescope has enabled blind pulsation searches, those without prior
knowledge of a radio timing ephemeris.  The LAT has discovered over 20
pulsars in blind searches of the gamma-ray data alone, and only a few
have since had radio confirmation.

The improved pulse-phase spectroscopy enabled by {\it Fermi} has
clarified the probable locale of the gamma-ray emission region. The two
major competing models for the site of this emission are the polar cap
(PC) near the stellar surface (e.g. \cite{dh82,dh96}) and the outer gap
(OG) near the light cylinder (e.g. \cite{chr86,Romani96}).
Super-exponential turnovers due to magnetic pair creation are expected
for PC models at the maximum photon energies in the GeV band, as opposed
to exponential shapes for OG models. Discrimination by {\it Fermi}-LAT
between these two shapes was anticipated prior to launch \cite{rh07}. 
In the vast majority of LAT pulsar spectra, the turnovers are
exponential, thereby favoring the OG scenario \cite{Abdo10_LATCAT}.
Furthermore, the separation of peaks in pulsars with more than one pulse
peak is generally of the order of 0.3--0.5 in phase, indicating a broad
radiation cone that is consistent only with high altitude emission
locales for dipolar fields (e.g. \cite{Watters09}). This evidence has
ushered in a new era for the $\gamma$-ray pulsar paradigm. Here, some
perspectives pertaining to this new era are offered, focusing first on
lower bounds to the emission radius from magnetic pair creation physics.
In addition, the GeV-band turnover energies \teq{\Emax} are assessed in
terms of the canonical radiation-reaction-limited curvature emission
picture, touching upon population characteristics for $\gamma$-ray pulsars.

\section{Magnetic Pair Creation Bounds for the Minimum Altitude of Emission}
 \label{sec:pprod_emax}

The emergence of gamma-rays from pulsars is probably intimately coupled
to the generation of electron-positron pairs in their magnetospheres.
Such a connection between radiation emission and pair creation was
posited in early radio-pulsar pictures \cite{Sturr71,rs75}. In the inner
magnetosphere of young pulsars, the dominant pair production process
involves a single photon interacting with the {\bf B} field,
\teq{\gamma\to e^+e^-}, with its non-magnetic, quantum counterpart
\teq{\gamma\gamma\to e^+e^-} only becoming competitive at high
altitudes.  Hence, \teq{\gamma\to e^+e^-} is the preserve of {\it polar
cap} (PC; \cite{dh82}) and {\it slot gap} (SG; \cite{mh03}) gamma-ray
pulsar models. It is a first-order QED process that is kinematically
forbidden in field-free regions, but can take place in an external
magnetic field, which can absorb momentum perpendicular to {\bf B}.  Its
generic physics properties are summarized in \cite{Baring08}. Due to
energy conservation, its rate \teq{R^{\rm pp}} possesses an absolute
threshold energy, \teq{\erg_{\gamma}\geq 2/\sin\theta_{\rm kB}}. Here
\teq{\theta_{\rm kB}} is the angle the photon momentum vector {\bf k}
makes with the magnetic field {\bf B}, and the photon energy
\teq{\erg_{\gamma}} is in units of \teq{m_ec^2}, a convention adopted
hereafter.

Above threshold, \teq{R^{\rm pp}}, which is averaged here over photon
polarizations, exhibits a large number of resonances, corresponding to
thresholds for the creation of pairs in excited states; these aggregate
to exhibit a characteristic sawtooth structure (e.g.
\cite{dh83,Baring08}). In pulsar contexts, considerable ranges of field
strengths and photon angles \teq{\theta_{\rm kB}} are sampled during
magnetospheric propagation: the resulting convolution smears out the
sawtooth appearance into a continuum. In this regime, the dependence on
\teq{\erg_{\gamma}} can be approximated by a compact asymptotic
expression (e.g. \cite{erber66}; and references in \cite{Baring08}):
\begin{equation}
   R^{\rm pp} \;\approx\; \dover{3}{8}
    \sqrt{\dover{3}{2}}\, \dover{\fsc c}{\lambar} \, \dover{B}{B_{\rm cr}}\, \sin\theta_{\rm kB}
  \;\exp \biggl\{ -\dover{8}{3\Upsilon} \biggr\}  \, ,\quad 
  \Upsilon \,\equiv\, \dover{B}{B_{\rm cr}}\, \erg_{\gamma} \sin\theta_{\rm kB} \,\ll\, 1\; ,
   \label{eq:pp_asymp_rate}
\end{equation}
where \teq{\Upsilon} is the critical asymptotic expansion parameter.
Here, \teq{\fsc =e^2/(\hbar c)} is the fine structure constant,
\teq{\lambar =\hbar/(m_ec)=3.862\times 10^{-11}}cm is the electron
Compton wavelength over \teq{2\pi}, and \teq{B_{\rm cr}=m_e^2c^3/(e\hbar
)=4.413 \times 10^{13}}Gauss is the quantum critical field, where the
cyclotron energy equals \teq{m_ec^2}. Near threshold, this asymptotic
result is an imprecise description of \teq{R^{\rm pp}} \cite{dh83}, and
can be improved considerably by careful treatment of the associated
mildly-relativistic regimes for the produced pairs (e.g.
\cite{Baring88}).

In polar cap pulsar models \cite{Sturr71,rs75,dh82}, primary curvature
photon emission is emitted at very small angles to the magnetic field,
well below pair threshold (e.g. \cite{bh01}). These photons will convert
into pairs only after traveling a distance \teq{s} that is a fraction of
the field line radius of curvature \teq{\rho_c}.  Above the polar cap,
the radius of field curvature is \teq{\rho_c = [Prc/2\pi]^{1/2}} for a
pulsar period \teq{P}, and exceeds the neutron star radius \teq{\rns}.
Pair creation ensues for small angles such that \teq{\sin\theta_{\rm kB}
\sim s/\rho_c}, and the argument \teq{\Upsilon} of the exponential
becomes a fraction of unity, i.e., when \teq{\erg_{\gamma}
B\sin\theta_{\rm kB} \gtrsim 0.2\, B_{\rm cr}}.  Accordingly, in young
pulsars, pair production will usually occur somewhat or well above
threshold when \teq{B \ll 0.1\, B_{\rm cr}}, and the attenuation mean
free path \teq{\lambda_{\rm pp} \sim \rho_c/\erg\, \max \{ 2, \;
0.2B_{\rm cr}/B \} } will then be much less than \teq{\rns}.
Consequently, the optical depth \teq{\tau_{\gamma}(\erg_{\gamma})} for
one-photon pair opacity will be a {\it rapidly-varying function} of
\teq{\erg_{\gamma}}, producing a dramatic attenuation signature at the
escape energy \teq{\eesc}, which typically lies above 100 MeV (see
\cite{bh01}).

The upshot is that attenuation by \teq{\gamma\to e^+e^-} should impose a
{\it super-exponential} cutoff in pulsar $\gamma$-ray spectra of the
approximate form \teq{dn_{\gamma}/d\erg_{\gamma}\propto \exp \{ -\alpha
\exp [-\eesc /\erg_{\gamma} ]\} } for some constant \teq{\alpha}.  If
\teq{\eesc\sim\emax\equiv \Emax/m_ec^2}, the severity of the cutoff
should produce \cite{dh82,dh96} a sharp turnover that is distinguishable
from those generated by other mechanisms.  It was widely anticipated
\cite{rh07} that the {\it Fermi}-LAT instrument would discern the
precise shape of such cutoffs in Vela and other pulsars, and accordingly
discriminate between the polar cap and outer gap models for their high
energy emission.  It has in fact done so. Early on in the first year of
observations by {\it Fermi}, the high count rates for Vela permitted the
exclusion of super-exponential turnovers at a high level of significance
\cite{Abdo09_Vela}.  This has become possible for the pool of 39 young
pulsars listed in the {\it Fermi}-LAT Pulsar Catalog in
\cite{Abdo10_LATCAT}, all of which reveal significant turnovers in the
1-10 GeV band consistent with simple exponentials. This clearly
demonstrates that \teq{\gamma\to e^+e^-} is not producing the observed
turnovers, proving that \teq{\tau_{\gamma}(\erg_{\gamma})<1} for
\teq{\erg_{\gamma} \lesssim \emax}.

This undeniable absence of the signature of active magnetic pair
creation in gamma-ray pulsar signals can be inverted to provide robust
lower bounds to emission altitudes, a protocol that has been adopted in
a number of {\it Fermi}-LAT pulsar papers.  Defining a marginal magnetic
pair opacity criterion \teq{\tau_{\gamma}(\emax )\approx 1} near the
pulsar surface, rearrangement yields an approximate dependence of pair
creation cutoff energies \teq{\emax} on \teq{B_0}, \teq{R_0} and pulsar
period \teq{P} (in seconds).  Setting \teq{\Emax \equiv \emax m_ec^2},
this can be summarized in a relation \cite{Baring04} that corresponds to
near-surface emission (\teq{r\lesssim 2\rns}):
\begin{equation} 
   \Emax \sim 0.4 \sqrt{P} \, \biggl( \dover{r}{\rns} \biggr)^{1/2} \; 
   \max \Biggl\{ 1,\; \dover{0.1\, B_{\rm cr}}{B_0}\,  
   \biggl( \dover{r}{\rns} \biggr)^3 \Biggr\}\; \hbox{GeV} \;\; . 
 \label{eq:emax_PC} 
\end{equation} 
This encapsulates the character of accurate numerics derived from codes 
employed in \cite{hbg97} and \cite{bh01}, which include pronounced 
effects of general relativity on spacetime curvature, field enhancement 
and photon energy.  In flat spacetime, the altitudinal dependence is 
weaker, corresponding to \teq{\emax\propto (r/\rns )^{5/2}} 
(e.g. \cite{zh00,Lee10}) above the magnetic poles.


\begin{figure}
 \centerline{
  \includegraphics[width=.75\textwidth]{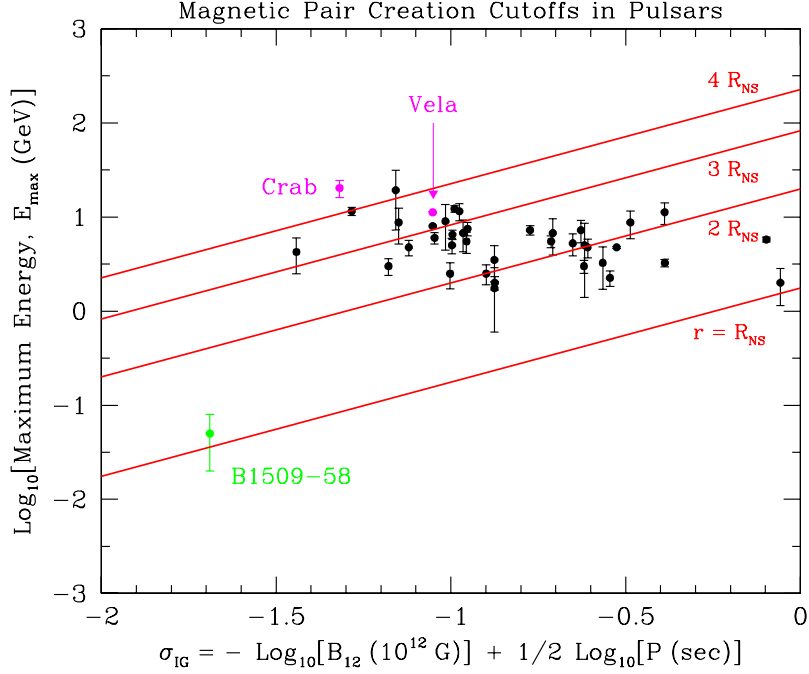}
  }
  \caption{The inferred maximum pulsar emission energies \teq{\Emax} 
as a function of the key polar cap (inner gap) magnetic pair creation 
attenuation parameter \teq{\sigmaIG} defined in Eq.~(\ref{eq:sigmaIG_def}).  
The spectral parameters used are the cutoff energies \teq{E_c} listed in Table~4
of \cite{{Abdo10_LATCAT}}, the First {\it Fermi}-LAT Catalog of Gamma-ray Pulsars.
For the 37 pulsars constituting the black points, \teq{\Emax =2.5E_c} was set;
for the Crab and Vela, \teq{\Emax=3.5E_c} (see text).  The high-field CGRO 
Comptel pulsar B1509-58 (J1513-5908) is also displayed.  No LAT millisecond 
pulsars are depicted.  The fiducial linear relationship in Eq.~(\ref{eq:sigmaIG_def}) 
is displayed as the diagonal lines for four different emission altitudes, 
\teq{r/\rns =1} (surface), \teq{2, 3, 4}, as labelled.
}
 \label{fig:hecutoff_PC}
\end{figure}

The pair opacity \teq{\emax} trend can be expressed alternatively as a
combination of observables appropriate for this inner gap (polar cap)
discussion.  The form is  
\begin{equation}
   \log_{10}\dover{\Emax}{\hbox{1 GeV}}\; \approx\; \sigmaIG 
        + \dover{7}{2}\, \log_{10}{\chiIG} +0.25
   \quad ,\quad
   \sigmaIG\; =\; - \log_{10}B_{12} + \dover{1}{2}\, \log_{10}P \quad .
 \label{eq:sigmaIG_def}
\end{equation}
which clearly encapsulates an anti-correlation with \teq{{\dot P}}, as
does Eq.~(\ref{eq:emax_PC}), since \teq{B_0 = 3.2\times 10^{19} (P\,
{\dot P})^{1/2}} \cite{MT77}.   Here \teq{\chiIG = r/\rns} is the
scaling of the radius appropriate for polar cap considerations. 
Figure~\ref{fig:hecutoff_PC} plots the phase space corresponding to
Eq.~(\ref{eq:sigmaIG_def}), using the measured spectroscopic cutoffs
\teq{E_c} for young {\it Fermi}-LAT pulsars in Table~4 of
\cite{Abdo10_LATCAT} to benchmark the maximum energies \teq{\Emax}.
Since these cutoffs are exponential in character, the action of magnetic
pair creation can be ruled out at energies below the window around
\teq{E_c}.  For this reason, a representative setting of \teq{\Emax =
2.5E_c} was deployed in the Figure for most of the pulsars, the set of
37 with points in black. This value can be adjusted on a case-by-case
basis, depending on the photon counting statistics.  Accordingly, the
alternative value of \teq{\Emax =3.5E_c} was adopted for the Crab and
Vela pulsars. Motivations for these exceptions are as follows: the Crab,
possesses high energy emission reflected in its pulsed detection by
MAGIC \cite{Aliu08_MAGIC_Crab}, and Vela, because its impressive count
rate as the brightest LAT-band pulsar in the sky permits extraordinary
spectroscopic determination of its turnover \cite{Abdo09_Vela}. The
uncertainties in \teq{E_c} were scaled similarly, yielding the
\teq{\Emax} error bars in the Figure.  PSR B1509-58, the young,
high-field pulsar detected by Comptel on CGRO, but not by EGRET, was
also placed in the Figure using an inferred \teq{\Emax} value from the
spectral results displayed in \cite{hbg97}.  It is the only gamma-ray
pulsar that might be consistent with \teq{\gamma\to e^+e^-} attenuation
operating near the stellar surface. Observe that no millisecond pulsars
(with \teq{\sigmaIG < -2}) are exhibited in Fig.~\ref{fig:hecutoff_PC}
as their surface fields are so low that \teq{\gamma\to e^{\pm}} pair
creation is exponentially improbable in their magnetospheres. The phase
space plot in Figure~\ref{eq:sigmaIG_def} clearly demonstrates that
there is no correlation between \teq{\emax} and \teq{\sigmaIG} or
\teq{{\dot P}} in the LAT pulsar population.

\section{Curvature Radiation at High Altitudes}
 \label{sec:curvature}

Given the above discussion it is evident that  the {\it Fermi} GeV-band
spectral cutoffs in young pulsars are probably related to the primary
emission mechanism.  Dating from the early models of \cite{Sturr71},
this mechanism is postulated to be {\it curvature radiation}, the
elementary dipole radiation resulting from pairs accelerating in the
curved magnetospheric fields.  The formalism for the curvature radiation
emissivity is identical to that for classical synchrotron radiation
\cite{Jack75,RL79}, but with the electron Larmor radius
\teq{r_{\hbox{\fiverm L}}=\gamma_e m_ec^2/(eB)} being replaced by the
radius of field curvature, \teq{\rho_c}.  Accordingly, the
characteristic dimensionless (i.e., in units of \teq{m_ec^2}) photon
energy scales for curvature radiation, \teq{\erg_c}, and synchrotron
emission \teq{\erg_s}, are
\begin{equation}
   \erg_c\; =\; \dover{3}{2}\,\gamma_e^3\, \dover{\lambar}{\rho_c}
      \quad ,\quad 
   \erg_s\; =\; \dover{3}{2}\,\gamma_e^3\, \dover{\lambar}{\rL}\,\sin\theta_e
   \; \equiv\; \dover{3}{2}\; \dover{B}{B_{\rm cr}}\,\gamma_e^2\,\sin\theta_e\quad .
  \label{eq:erg_c_curv_synch}
\end{equation}
For curvature emission to seed pulsar gamma-ray signals, \teq{\erg_c}
must exceed a few GeV in the magnetosphere; this is routinely satisfied
in polar cap (e.g. \cite{dh96}), slot gap \cite{mh03} and outer gap
(e.g. \cite{Romani96}) models, for both young pulsars and millisecond ones.

At its characteristic photon energy, the differential emissivity per
electron scales as \teq{\fsc c/\lambar\, \gamma_e^{-2}} (e.g. see
Eq.~(2) of \cite{bb04}) for both the curvature and synchrotron
processes.  When this emissivity scale is multiplied by the square of
the photon energy scale in Eq.~(\ref{eq:erg_c_curv_synch}), the
corresponding magnitudes for the curvature radiation and synchrotron
cooling rates result; in full form these rates are
\begin{equation}
   \gammadotCR\; =\; -\dover{2}{3}\, \dover{r_0 c}{\rho_c^2}\;\gamma_e^4
   \quad ,\quad 
   \gammadotSR\; =\; -\dover{2}{3}\, \dover{r_0 c}{r_g^2}\;\gamma_e^2\,\sin^2\theta_e\quad .
 \label{eq:CR_SR_coolrate}
\end{equation}
Here \teq{r_g=m_ec^2/(eB)=\lambar B_{\rm cr}/B} is \teq{c} divided by
the cyclotron frequency, and \teq{r_0=\fsc \lambar =e^2/(m_ec^3)} is the
classical electron radius. The precise numerical factors can be derived
using the synchrotron formulation of \cite{RL79}; for the curvature
process this requires the specialization to pitch angles
\teq{\theta_e=\pi /2}.  From Eq.~(\ref{eq:CR_SR_coolrate}) one derives
the criterion for the {\it dominance of a synchrotron signal over a
curvature one}: \teq{\gamma_e \lesssim\; \rho_c\,\sin\theta_e/r_g}.
While \teq{\theta_e\approx 0} is presumed for the primary emission, pair
creation usually generates secondary photons at significant angles to
the field, so that this inequality is almost always satisfied:
synchrotron photons therefore abound in subsequent generations of a
cascade. Since \teq{\rho_c/r_g\gg 10^{15}} near the polar cap in young
pulsars, only very small \teq{\theta_e} are required to spawn the
swamping of primary curvature emission by a synchrotron component in the
inner magnetosphere.  This possibility should be borne in mind for
future refinements of the gamma-ray pulsar paradigm.

The maximum Lorentz factor of the electrons emitting in gamma-ray
pulsars is a function of the accelerating potential adopted in a given
model. If the electric field component along the local magnetic field is
\teq{E_{\parallel}} (presumed positive), then the energy gain rate in
the electrostatic gap is \teq{\gammadotacc m_ec^2 = eE_{\parallel}c},
i.e. the electron speed \teq{c} times the force.  The maximum possible
electron Lorentz factor \teq{\gammax} is realized when this gain equals
the loss rate due to radiative cooling, which is the {\it
radiation-reaction limit} generally ascribed to curvature emission in
gamma-ray pulsar models.  Setting \teq{-eE_{\parallel}/(m_ec)} equal to
the first result in Eq.~(\ref{eq:CR_SR_coolrate}) yields a maximum
\teq{e^-} Lorentz factor \teq{\gammax\sim [3\rho_c^2 E_{\parallel}/(2 e)
]^{1/4}}, which can then be inserted into
Eq.~(\ref{eq:erg_c_curv_synch}) to specify the approximate maximum
photon energy (in units of \teq{m_ec^2}) for curvature radiation
reaction-limited acceleration:
\begin{equation}
   \emax\; =\; \left(\dover{3}{2}\right)^{7/4}\, \lambar\,\rho_c^{1/2}\,
      \left( \dover{E_{\parallel}}{e} \right)^{3/4} \quad .
 \label{eq:emax_CR}
\end{equation}
This form for \teq{\emax} is routinely cited in {\it Fermi}-LAT
publications on gamma-ray pulsars (e.g. \cite{Abdo10_LATCAT}). In
principal, the observed cutoff energy can be less than this value if
attenuation by pair production, magnetic or two-photon, is efficient, or
the acceleration is terminated by mechanisms other than curvature
cooling.  Hence, the observed GeV-band turnovers in dozens of young
pulsars in the First {\it Fermi}-LAT Pulsar Catalog provide lower bounds
to the accelerating potential, modulo the altitude of emission.

A natural, fiducial scale for the parallel electric field can be
obtained from Lorentz transformations associated with a rotating
magnetosphere.  This establishes \teq{r\vec{\Omega}\times\vec{B}/c} as
the co-rotational component of \teq{\vec E}, providing the ideal MHD
contribution. Yet within the plasmasphere, a non-co-rotational, parallel
component \teq{E_{\parallel}} emerges due to departures from
Goldreich-Julian current flow (e.g. \cite{Shibata95,Takata06}): it is
putatively of the same order of scaling \teq{\sim r\Omega B/c}.  This is
the most optimistic scenario, being adopted for example in the seminal
outer gap pulsar model of \cite{chr86}, and leads to
\teq{E_{\parallel}\sim e/(r_0\lambar ) \chi^{-2}\, \Blc/B_{\rm cr}} for
\teq{\chi =r/\rlc} describing the scaling of the acceleration/emission
radius in terms of the light cylinder radius \teq{\rlc =Pc/(2\pi )} that
will be employed hereafter. These define so-called thick outer gaps
(e.g. \cite{zc97}). Here \teq{\Blc} is the field strength at
\teq{r=\rlc}, essentially along the last open field line. Note that the 
dependence of \teq{E_{\parallel}}, \teq{\Blc} and the radius of
curvature on the obliquity of the rotator, and colatitude \teq{\Theta},
is omitted in this discussion. Using the scaled radius of curvature
\teq{\nu_c =\rho_c/(\chi\rlc )}, one can then insert this nominal value
for \teq{E_{\parallel}} into Eq.~(\ref{eq:emax_CR}). This can be cast in
terms of observables \teq{P} and \teq{{\dot P}}.  For a surface polar
field \teq{B_0=3.2\times 10^{19} (P\, {\dot P})^{1/2}\equiv
10^{12}B_{12}}Gauss and a neutron star radius of \teq{\rns =10^6}cm,
since \teq{\Blc = B_0\, (\rns /\rlc)^3}, setting \teq{P=0.1P_{-1}}sec
yields a result of
\begin{equation}
   \Emax\;\approx\; \dover{8.0\nu_c^{1/2}}{{\chi}}
   \, (\epspar\, B_{12})^{3/4}\, (P_{-1})^{-7/4}\; \hbox{GeV}   \quad  .
 \label{eq:Emax_CR_fid}
\end{equation}
This scaling is commensurate with that obtained in Eq.~(4) of
\cite{Romani96}.  Clearly \teq{\Emax} rises monotonically as the
acceleration locale moves from the light cylinder towards the neutron
star surface, due to the rapidly rising inductive \teq{B}-field.  An
additional {\it electrostatic decrement} factor \teq{\epspar \leq 1} has
been introduced in Eq.~(\ref{eq:Emax_CR_fid}) to facilitate the
subsequent discussion.  It is an acceleration efficiency factor,
representing the departure of \teq{E_{\parallel}} below the fiducial,
optimistic value of \teq{E_{\parallel}\sim r\Omega B/c};  in general it
is a function of the altitude parameter \teq{\chi}, the colatitude
\teq{\Theta} and the rotator obliquity angle \teq{\alpha}.


\begin{figure}
 \centerline{
  \includegraphics[width=.75\textwidth]{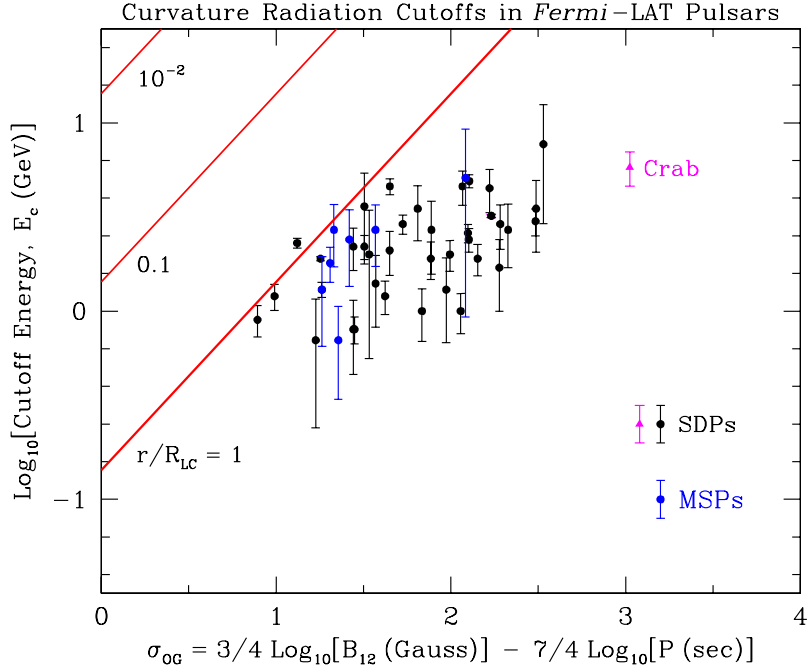}
  }
  \caption{Observed {\it Fermi}-LAT pulsar emission cutoff energies 
as a function of the key outer gap curvature radiation reaction-limited 
acceleration parameter \teq{\sigmaOG} defined in Eq.~(\ref{eq:sigmaOG_def}).  
The data come from the First {\it Fermi}-LAT Pulsar Catalog \cite{Abdo10_LATCAT}, 
with classic spinning-down pulsars (SDPs) in black, and millisecond pulsars (MSPs) 
in blue, as indicated.  The Crab pulsar point from the Fermi dataset is highlighted, noting that 
its \teq{E_c} from the MAGIC detection \cite{Aliu08_MAGIC_Crab} is considerably higher.
The diagonal lines labelled with \teq{\chi = r/\rlc} values constitute solutions of  
Eq.~(\ref{eq:sigmaOG_def}) for \teq{\epspar=1} and \teq{\nu_c=1}.
}
 \label{fig:hecutoff_OG}
\end{figure}

To interpret this outer magnetospheric curvature radiation cutoff
scaling, Figure~\ref{fig:hecutoff_OG} displays the phase space
corresponding to Eq.~(\ref{eq:Emax_CR_fid}), plotting \teq{\log_{10}E_c}
versus the appropriate combination of observables \teq{P} and \teq{{\dot
P}}; in this graph, the cutoff energy \teq{E_c} serves as a proxy for
\teq{\Emax}.   For the curvature radiation expectation in an outer gap
(OG) near the light cylinder, this is (for \teq{{\hat \chi} =
\chi\,\nu_c^{-1/2}\epspar^{-3/4}})
\begin{equation}
   \log_{10}\dover{\Emax}{\hbox{1 GeV}}\; \approx\; \sigmaOG - \log_{10}{\hat \chi} -0.85
   \quad ,\quad
   \sigmaOG\; =\; \dover{3}{4}\, \log_{10}B_{12} - \dover{7}{4}\, \log_{10}P \quad .
 \label{eq:sigmaOG_def}
\end{equation}
The graph is derived from the LAT pulsar data in Tables~1 and~4 of
\cite{Abdo10_LATCAT}.  Since the \teq{\Emax} values appear in a
relatively narrow range, with considerable spread in \teq{\sigmaOG},
there appears to be no clear linear correlation, just the
suggestion of one. The diagonal lines corresponding to \teq{{\hat \chi}
=1, 0.1, 10^{-2}} in the Figure provide an approximate guide to read off
the scaled altitude.  It is clear that this case of maximal
\teq{E_{\parallel}} cannot quite accommodate the turnover energies of
the LAT pulsars being attributed to cooling-limited curvature emission
within the light cylinder: the observations mandate that the
electrostatic decrement lie in the range \teq{0.05\lesssim
\epspar\lesssim 1} if \teq{\chi=1}. This essentially implies
acceleration potentials somewhat weaker than the co-rotational benchmark
\teq{r\Omega B/c}, or considerably so (i.e., \teq{\epspar\lesssim
10^{-2}}) if the outer gap emission altitude lies well within the light
cylinder radius, i.e. \teq{\chi\lesssim 0.1}.  This observational
diagnostic becomes less constraining if some two-photon pair attenuation
is active in the emission region, thereby lowering the observed cutoff
energy \teq{E_c} below the curvature radiation \teq{\Emax}.

The phase space plot in Fig.~\ref{fig:hecutoff_OG} exhibits a remarkable
property --- that the youngish, dipole-torque spin-down pulsars (SDPs)
and old, recycled millisecond pulsars (MSPs) occupy the same locale. 
The $x$-coordinate, \teq{\sigmaOG}, is an approximate linear function of
\teq{\log_{10}\Blc}, and the field \teq{\Blc} is fairly similar for the
two populations (see Table 1 of \cite{Abdo10_LATCAT}, and compare
Fig.~\ref{fig:hecutoff_OG} here with their Figure~7 that displays the
array of {\it Fermi}-LAT pulsar \teq{\emax} values versus
\teq{\log_{10}\Blc}). This approximate coincidence of \teq{\Blc} values
in SDPs and MSPs was, of course, well-known for the radio pulsar
population. Yet, the clustering of the gamma-ray \teq{\Emax} around
similar values for SDPs and MSPs is a new insight enabled by {\it
Fermi}.  It is strongly suggestive that similar physical acceleration
locales and radiative emission mechanisms operate in the two
populations.

\vskip 5pt\noindent
{\bf Concluding Remarks:} The picture of high altitudes for $\gamma$-ray
emission in young pulsars is now well-established.  Yet questions still
remain. Is the primary mechanism curvature emission, or can small-angle
synchrotron radiation contribute significantly to the signal?  What is
the role of two-photon pair creation in controlling the range of
observed \teq{\emax}?  Does magnetic pair production at low altitudes
still enhance the multiplicity of pairs being pumped into surrounding
pulsar wind nebulae? These issues will capture the focus of theorists in
the years to come as the {\it Fermi} pulsar legacy further unfolds.

\bibliographystyle{aipproc}

\begin{thebibliography}{12}

\bibitem{Abdo09_Vela}
Abdo, A., et al. 2009, \apj,\vol{696}{1084}
\bibitem{Abdo10_LATCAT}
Abdo, A., et al. 2010, \apjsupp,\vol{187}{460}
\bibitem{Aliu08_MAGIC_Crab}
Aliu, E., et al. 2008, Science, \vol{322}{1221}
\bibitem{Baring88}
Baring, M.~G. 1988, \mnras,\vol{235}{51}
\bibitem{Baring04}
Baring, M.~G. 2004, \asr,\vol{33}{552}
\bibitem{Baring08}
Baring, M.~G. 2008, in Proc. CASYS '07 Conference
     {\it Computing Anticipatory Systems},
     ed. D. M. Dubois (AIP Conf. Proc. 1051, New York), p.~53.
\bibitem{bb04}
Baring, M.~G.  \& Braby, M.~L. 2004, \apj,\vol{613}{460}
\bibitem{bh01}
Baring, M.~G. \& Harding A.~K. 2001, \apj,\vol{547}{929}
\bibitem{chr86}
Cheng K. S., Ho C. \& Ruderman M. 1986, \apj,\vol{300}{500}
\bibitem{dh82}
Daugherty, J.~K. \& A.~K. Harding 1982, \apj,\vol{252}{337}
\bibitem{dh83}
Daugherty, J.~K. \& Harding A.~K. 1983, \apj,\vol{273}{761}
\bibitem{dh96}
Daugherty, J.~K. \& Harding A.~K. 1996, \apj,\vol{458}{278}
\bibitem{erber66}
Erber, T. 1966, \rmp,\vol{38}{626}
\bibitem{hbg97}
Harding A.~K., Baring, M.~G. \& Gonthier, P.~L. 1997, \apj,\vol{476}{246}
\bibitem{Jack75}
Jackson, J.~D. 1975, {\it Classical Electrodynamics}
   (Wiley and Sons, New York)
\bibitem{Lee10}
Lee, K.~J., et al. 2010, \mnras,\vol{405}{2103}
\bibitem{MT77}
Manchester, R.~N. \& Taylor, J.~H. 1977, {\it Pulsars} (Freeman, San Francisco).
\bibitem{mh03}
Muslimov, A.~G. \& A.~K. Harding 2003, \apj,\vol{588}{430}
\bibitem{rh07}
Razzano, M. \& Harding, A.~K. 2007, in {\it The First GLAST Symposium},
   eds. S. Ritz, P.~F. Michelson \& C. Meegan (AIP Conf. Proc. 921) p.~413.
\bibitem{Romani96}
Romani, R.~W. 1996,  \apj,\vol{470}{469}
\bibitem{rs75}
Ruderman, M.~A. \& Sutherland, P.~G. 1975, \apj,\vol{196}{51}
\bibitem{RL79}
Rybicki, G. \& Lightman, A. 1979, {\it Radiative Processes in
   Astrophysics,} (Wiley \& Sons, New York)
\bibitem{Shibata95}
Shibata S. 1995, \mnras,\vol{276}{537}
\bibitem{Sturr71}
Sturrock, P.~A. 1971, \apj,\vol{164}{529}
\bibitem{Takata06}
Takata, J., Shibata, S., Hirotani, K. \& Chang, H-K. 2006, \mnras,\vol{366}{1310}
\bibitem{Watters09}
Watters, K.~P., Romani, R.~W., Weltevrede, P. \& Johnston, S. 2009, \apj,\vol{695}{1289}
\bibitem{zh00}
Zhang, B. \& Harding, A.~K. 2000, \apj,\vol{532}{1150}
\bibitem{zc97}
Zhang L. \& Cheng K. S. 1997, \apj,\vol{487}{370}

\end{thebibliography}

\end{document}